\documentstyle[aps,epsf,floats,times]{revtex}
\begin{document}
\draft

\wideabs{

\title{Dual-resonator speed meter for a free test mass
}

\author{Vladimir B.\ Braginsky$^1$, Mikhail L.\ Gorodetsky$^1$,
Farid Ya.\ Khalili$^1$, and Kip S.\ Thorne$^2$}

\address{$^1$Physics Faculty, Moscow State University, Moscow Russia}
\address{$^2$Theoretical Astrophysics, California Institute of
Technology, Pasadena, CA 91125}

\date{Received 24 June 1999}

\maketitle

\begin{abstract}

A description and analysis are given of a ``speed meter'' for monitoring a
classical force that acts on a test mass.  This speed meter is based on two
microwave resonators (``dual resonators''),
one of which couples evanescently to the position of
the test mass.  The sloshing of the resulting signal between the resonators,
and a wise choice of where to place the resonators' output waveguide, produce
a signal in the waveguide that (for sufficiently low frequencies) is
proportional to the test-mass velocity (speed) rather than its position.  This
permits the speed meter to achieve force-measurement sensitivities better
than the standard  quantum limit (SQL), both when operating in a narrow-band
mode and a wide-band mode.  A scrutiny of experimental issues shows that it
is feasible,  with current technology, to construct a demonstration  speed
meter that beats the wide-band SQL by a factor 2.  A concept is sketched for
an adaptation of this speed meter to optical frequencies; this adaptation
forms the basis for a possible LIGO-III interferometer that could beat the
gravitational-wave standard quantum limit $h_{\rm SQL}$, but  perhaps only by
a factor 
$1/\xi = h_{\rm SQL}/h \alt 3$ 
(constrained by losses in the optics) and at the price of a very high
circulating optical power --- larger by $\xi^{-2}$ than that  required to
reach the SQL.

\end{abstract}

}

\narrowtext

\section{Introduction}
\label{sec:intro}
A conceptual design for a {\it quantum speed meter} was
proposed several years ago \cite{speedmeter}.
This speed meter couples to
the velocity of a free test mass and thereby can monitor a classical force
that acts on the test mass with a precision better than the Standard Quantum
Limit (SQL).

The motivation for coupling to test-mass velocity rather than position is
that (in the absence of the coupling) the test-mass velocity is equal to
momentum divided by mass; and momentum, by contrast with position, is a
constant of the test mass's free motion, so it commutes with itself at
different times and is a quantum nondemolition (QND) observable
\cite{QuantumMeasurement}.  This enables the speed meter to beat the
classical-force SQL without any special squeezed-state preparation of the
speed meter's microwave pump field or frequency-dependent homodyne detection
of its output signal field.  By contrast, to beat the classical-force SQL, a
meter that couples to position must incorporate a squeezed-state pump and/or
frequency-dependent homodyne detection; see Appendix \ref{app:comparison}.

In Ref.\ \cite{speedmeter} two variants of the speed meter were suggested,
one based on an optical-fiber delay line and the other on coupled microwave
resonators
(``dual resonators'').  In this paper we analyze in detail the
dual-resonator scheme
and show that it can be realized in principle with current experimental
technology.

An important possible application of this speed meter is as the readout
device for a new class of laser-interferometer gravitational-wave antennas
that may beat the SQL while using unusually low laser power
\cite{NL96,optbar,symph}.

\begin{figure}
\unitlength=2pt

\begin{picture}(180,90)

\thicklines      
\put(45,55){\line(1,0){30}}\put(75,55){\line(0,1){30}}
\put(45,55){\line(0,1){30}}\put(45,85){\line(1,0){30}}
\put(80,85){\makebox(0,0)[cc]{\large{2}}}

\thinlines       
\put(15,72){\line(1,0){30}}\put(45,70){\oval(4,4)[r]}
\put(25,72){\vector(1,0){10}}
\put(30,70){\makebox(0,0)[ct]{$U_0\cos\omega_e t$}}

\thicklines      
\put(75,10){\line(0,1){35}}
\put(45,10){\line(0,1){35}}\put(45,45){\line(1,0){30}}
\put(80,45){\makebox(0,0)[cc]{\large{1}}}

\thicklines      
\put(46,10){\line(1,0){28}}\put(74,10){\line(0,1){5}}
\put(46,10){\line(0,1){5}}\put(46,15){\line(1,0){28}}
\put(46,10){\makebox(28,5)[cc]{\large{m}}}

\thinlines
\put(50,30){\vector(0,-1){15}}\put(50,30){\vector(0,1){15}}
\put(52,30){\makebox(28,5)[lc]{\large{d}}}

\thinlines       
\put(75,32){\line(1,0){35}}\put(75,30){\oval(4,4)[l]}
\put(77,36){\vector(1,0){10}}
\put(90,36){\makebox(0,0)[lc]{$U^{out}(t)$}}
\put(87,28){\vector(-1,0){10}}
\put(89,28){\makebox(0,0)[lc]{$U_e(t)$}}

\thinlines       
\put(62,45){\line(0,1){10}}
\put(60,45){\oval(4,4)[b]}
\put(60,55){\oval(4,4)[t]}
\put(64,50){\makebox(0,0)[lc]{$\Omega$}}

\end{picture}

\caption{Schematic diagram of the coupled-resonator quantum speed meter.}
\label{fig:speedmeter}
\end{figure}
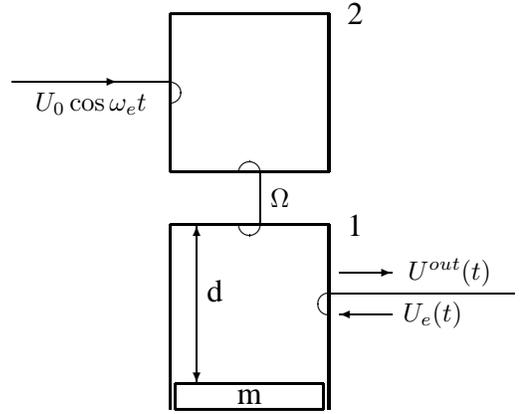

The speed meter proposed in Ref.\ \cite{speedmeter} is based on two
identical, weakly coupled microwave resonators as shown in Fig.\
\ref{fig:speedmeter}. It is a fascinating characteristic of such coupled
resonators that, when one is driven at their common eigenfrequency
$\omega_e$, it is the other that becomes excited.  Resonator 2 is pumped on
resonance by the voltage $U_0\cos\omega_e t$ of an input waveguide, so
resonator 1 becomes excited at frequency $\omega_e$.  The eigenfrequency of
resonator 1 is modulated by the position $x$ of the test mass

\begin{equation}
  \tilde\omega_e(x) = \omega_e\left(1-\frac{x}{d}\right),
\end{equation}
where $d$ is a length that characterizes the resonator's tunability (cf.\
Sec.\ \ref{sec:realization}); this modulation puts a voltage signal
proportional to position $x$ into resonator 2,
and a voltage signal proportional
to velocity $dx/dt$ into resonator 1.  The velocity signal flows from
resonator 1 into an output waveguide, from which it is monitored.

One can understand the production of this velocity signal as follows:
The weak coupling between the resonators causes voltage signals to
periodically slosh
from one resonator to the other at the
frequency $\Omega \ll \omega_e$.  After each cycle of
sloshing,
the sign of the signal is reversed, so the net signal in resonator 1 is
proportional to the difference of the position at times $t$ and
$t+2\pi/\Omega$, i.e.\ is proportional to the test-mass velocity so long
as the the test mass's frequencies $\omega$ of oscillations are $\omega \ll
\Omega$.

Actually, we shall find that the optimal regime of operation for the speed
meter is at signal frequencies $\omega \sim \Omega$.  In this regime, the
voltage signal in resonator 1 and the corresponding output voltage signal are
sums over time derivatives of the test-mass position [Eqs.\
(\ref{C1}--\ref{C2})]. Correspondingly, the speed meter does not monitor just
the speed, but rather the speed plus time derivatives of the speed.

In this paper we shall analyze in detail the operation of the speed meter,
first ignoring the resonators' dissipative losses and associated
noise (Sec.\ \ref{sec:no_losses}), then including the losses and noise
(Sec.\ \ref{sec:losses}).  We shall express the speed meter's performance in
terms of the spectral density $S(\omega)$ of the net noise that it
produces when monitoring a classical signal force $F_s(t)$ that is
acting on the test
mass.  As a foundation for this, in Sec.\ \ref{sec:SQL} we discuss the
SQL for force measurements in the language of spectral density.  In Sec.\
\ref{sec:realization}
we discuss the most serious practical impediments to achieving a sensitivity
that actually beats the SQL by a significant factor and conclude that
a demonstration experiment is feasible with current technology.  In
Appendix \ref{app:comparison}
we compare this speed meter with a position meter based on
a single microwave resonator with homodyne readout in the output waveguide
at a frequency-dependent homodyne phase (``quantum variational technique'');
and in Appendix \ref{app:GWAntenna}
we describe a speed-meter-based
conceptual design for a LIGO-type gravitational-wave antenna that can beat
the gravitational-wave SQL, but requires very high light power.

\section{Standard Quantum Limits}
\label{sec:SQL}

The standard quantum limits (SQL) for measurement of a classical
signal force $F_s(t)$ acting on
a free test mass, as usually given in the literature (e.g.\
\cite{QuantumMeasurement}), are not convenient since they are based on
some assumed shape of the force's time dependence (most commonly a single-cycle
sinusoid or a long, monochromatic wave train).  In this paper we prefer
the greater generality of a SQL expressed in terms of the 
two sided spectral density
$S(\omega)$ for the net noise in a measurement of $F_s(t)$; $S(\omega)$ is
defined such that
for optimal signal processing the 
measurement's power signal to noise ratio is 
\begin{equation}
{S\over N} = \int_{-\infty}^\infty
{|F_s(\omega)|^2\over S(\omega)} {d\omega\over2\pi}\;.
\label{SoverN}
\end{equation}
Here $F_s(\omega)$ is the Fourier transform of $F_s(t)$
\begin{equation}
F_s(t) = \int_{-\infty}^{\infty} F_s(\omega)
e^{-i \omega t} {d\omega\over2\pi}\;,
\label{FTConvention}
\end{equation}
in which we adopt
the $2\pi$ convention of signal processing theory and microwave technology,
and the $e^{-i\omega t}$ sign convention of quantum physics (so field
amplitudes and annihilation operators evolve
as $e^{-i\omega t}$ in the Heisenberg Picture).

\subsection{Wideband SQL}
\label{sec:widebandSQL}

An ordinary position meter (sometimes called coordinate meter)
monitors the position $x(t)$ of a free test mass, and thereby deduces
the classical signal force $F_s(t)$ that acts on the mass.
The spectral density of the net noise in this force monitoring is
\begin{equation}
  S(\omega)=m^2\omega^4 S_x + S_F\;,
\label{SSxSf}
\end{equation}
where
$m$ is the mass of the test mass,
$S_x(\omega)$ is the spectral density of the noise $x_{\rm m}(t)$
that the meter
superimposes on the output position signal, and $S_F(\omega)$ is
the spectral density of the fluctuating back-action force $F_{\rm BA}(t)$
that the meter
exerts on the test mass.  For an ordinary position meter, $x_{\rm m}(t)$ and
$F_{\rm BA}(t)$
are uncorrelated, and the Heisenberg uncertainty principle implies
that $S_x S_F
\ge \hbar^2/4$ \cite{QuantumMeasurement}.  We shall assume
that the position meter is as perfect as possible, corresponding to
equality in this uncertainty relation.  If the spectrum of
the classical force is concentrated near the frequency $\omega_F$ and
the position meter is optimally tuned for monitoring this force (so the
ratio $S_F/S_x$ is adjusted to make the two terms in
Eq.\ (\ref{SSxSf}) equal at $\omega=\omega_F$),
then the net spectral density is
\begin{eqnarray}
  S(\omega) &=& \hbar m\omega^2 \quad \mbox{ for } \omega = \omega_F\;,
\nonumber\\
  S(\omega) &\ge& \hbar m\omega^2 \quad \mbox{ for other values of } \omega\;.
\end{eqnarray}
This is the spectral-density form of the SQL.
The corresponding minimum detectable amplitude for a force that lasts for a
time $\tau_F$ is
\begin{equation}
  F_{\rm WB\ SQL} \simeq \sqrt\frac{\hbar m \omega_F^2 }{\tau_F}\;.
\label{WB_SQL}
\end{equation}
This is the usual form of the wide-band SQL for a sinusoidal force.

In order for the meter to beat this usual wide-band SQL by a factor
$\xi_{\rm WB} < 1$,
\begin{equation}
  F \simeq \xi_{\rm WB} F_{\rm WB\ SQL},
\end{equation}
the spectral density
of the net noise must obey the condition
\begin{equation}
  S(\omega) = \xi_{\rm WB}^2\hbar m\omega^2
\label{over_WB_SQL}
\label{xiWBdef}
\end{equation}
in the range of frequencies of the detected force.
We shall regard
Eq.\ (\ref{xiWBdef})
as a definition of the amount $\xi_{\rm WB}(\omega)$ by which
our speed meter beats the broad-band SQL.

\subsection{Narrowband SQL}
\label{sec:narrowbandSQL}

If the test mass has a restoring force so it is an oscillator
with eigenfrequency $\omega_0$, and/or the noises
$x_{\rm m}(t)$ and $F_{\rm BA}(t)$ are correlated (with cross spectral density
$S_{xF}$), then the net noise in the measurement of $F_s(t)$ is

\begin{equation}
S(\omega) = m^2(\omega^2-\omega_0^2)^2 S_x -2m(\omega^2-\omega_0^2)S_{xF}
+ S_F\;.
\label{Snet}
\end{equation}
For such a system, the noise can be made especially small in a narrow-band
measurement centered on the frequency

\begin{equation}
  \omega_{\rm meter}^2 = \omega_0^2 + \frac{S_{xF}}{mS_x}\;.
\end{equation}
If the noises $S_x$ and $S_F$ can be regarded as constant over that narrow
band, and they are constrained
only by the Heisenberg uncertainty relation $S_x S_F - S_{xF}^2 = \hbar^2/4$
\cite{QuantumMeasurement}), then
\begin{equation}
  S(\omega)=m^2(\omega_{\rm meter}^2-\omega^2)^2 S_x + \frac{\hbar^2}{4S_x}\;.
\label{S_oscill} \end{equation}

Suppose, now, that we use such a SQL-limited meter to measure a sinusoidal
force with frequency $\omega_F \simeq \omega_{\rm meter}$ and with duration
$\tau_F \gg 2\pi/\omega_F$ so the bandwidth of the force is
$\Delta\omega = 2\pi/\tau_F \ll \omega_F$.  Then, if $S_x$ is optimized,
the amplitude of the minimum detectable force [as computed by setting
$S/N\simeq 1$ in Eq.\ (\ref{SoverN})] is at the narrow-band SQL:
\begin{equation}
  F \simeq F_{\rm NB\ SQL} = \frac{1}{\tau_F}\sqrt{\hbar\omega_F m}\;.
  \label{NB_SQL}
\end{equation}

Correspondingly, in order to beat the narrow-band SQL,
the meter's net spectral density (\ref{Snet}) in the vicinity of some
frequency $\omega_{\rm meter}$ must have the form:
\begin{equation}
  S(\omega)=A(\omega_{\rm meter}^2-\omega^2)^2 + B \;,
\label{narrow_band_S}
\end{equation}
where the parameters  $A$ and $B$ (whose ratio is adjustable) satisfy
\begin{equation}
  AB = \xi_{\rm NB}^4\frac{\hbar^2 m^2}{4} \label{over_NB_SQL}\;.
\label{AB}
\end{equation}
The factor $\xi_{\rm NB}<1$ is the amount by which
the minimum detectable force is below the narrow-band SQL (\ref{NB_SQL}).

Another viewpoint on $\xi_{\rm NB}$ is the following:  Define
\begin{equation}
\bar S_F\equiv S_F-\frac{S_{xF}^2}{S_x}\;, \quad
\bar L_\omega \equiv L_\omega+\frac {S_{xF}}{S_x}\;,
\end{equation}
where $L_\omega$ is the spectral response of the test mass
($L_\omega=-m\omega^2$ for a free mass and $L_\omega=m(\omega_0^2-\omega^2)$
for a lossless oscillator). Then the net noise [Eq.\ (\ref{Snet})
with $m(\omega^2 -
\omega_0^2)$ replaced by $-L_\omega$] takes the form
\begin{equation}
S(\omega)=\bar L_\omega^2S_x+\bar S_F\;.
\end{equation}
If the noises are constrained only by the
Heisenberg uncertainty relation, $S_x S_F - S_{xF}^2 = S_x \bar S_F =
\hbar^2/4$, and one chooses $\omega_{\rm meter}$ to be at a zero of
$\bar L_\omega^2 $, then comparison with Eqs.\ (\ref{narrow_band_S}) and
(\ref{AB}) reveals the following expression for the amount by which the
narrow-band SQL can be beaten:
\begin{equation}
\xi^2_{NB}=\left|\frac{\bar L_\omega}{m(\omega^2_{meter}-\omega^2)}\right| \;.
\label{LW}
\end{equation}

\section{Microwave Speed Meter in the Lossless Limit}\label{sec:no_losses}

\subsection{Equations of motion and their solution}

When we neglect all losses in the test mass and in the resonators (and all
associated fluctuating forces), except those due to coupling to the output
waveguide, then the equations of
motion for the speedmeter of Fig.\ \ref{fig:speedmeter} take the following
form \cite{yurke}:

\begin{mathletters}
\label{eom}
\begin{eqnarray}
  &&\frac{d^2 q_1(t)}{dt^2} + 2\delta_e\frac{d q_1(t)}{dt} +
    \omega_e^2\left(1-\frac{x(t)}{d}\right)^2 q_1(t) \nonumber\\
  &&\quad =
    2\omega_e\Omega q_2(t) + 2\frac{\omega_e}{\rho}U_e(t)\;,
      \label{eomq1}\\
  &&\frac{d^2 q_2(t)}{dt^2} + \omega_e^2 q_2(t) =
    2\omega_e\Omega q_1(t) + 2 \frac{\omega_e}{\rho}U_0\cos\omega_e t\;,
      \label{emoq2}\\
  &&m{d^2 x\over dt^2} = {\rho\omega_e\over d}q_1^2
     - {\rho\omega_e\over 2d}q_0^2 + F_s(t)\;.
      \label{eomx}
\end{eqnarray}
\end{mathletters}
Here the notation is as follows:
\begin{itemize}
\item
$\omega_e$ is the common (angular)
eigenfrequency of the two resonators and $\Omega \ll \omega_e$ is
the weak-coupling frequency at which energy sloshes between the two
resonators;
\item
$q_{1,2}(t)$ are generalized coordinates of resonators 1 and 2,
so defined that the energy in resonator $j$ is
\begin{equation}
E_j = {\rho\over 2\omega_e}\dot q_j^2 + {\rho\omega_e \over 2}q_j^2\;,
\label{Ej}
\end{equation}
with an overdot representing a time derivative;
\item
$\rho$ is
the characteristic impedance of the resonators;
\item
$\delta_e \equiv 1/2\tau_e^*$ where $\tau_e^*$
is the relaxation time of resonator 1 due to energy
flowing into the output waveguide;
\item $U_e$ is the fluctuating voltage imposed on resonator 1
from the output waveguide;
\item $U_0$ is the driving voltage from the input waveguide, and is
assumed to be the result of a very strong waveguide field and a very weak
coupling to the resonator, so the
waveguide's fluctuational voltages can be ignored and $U_0$ can be regarded
as a classical c-number;
\item
$x(t)$ is the position of the free test mass, $d$ is the tuning length of
resonator 1, and $x/d$ is assumed to be so small that $(x/d)^2$ can be
neglected;
\item
$F_s (t)$ is the classical signal force acting on the test mass;
\item
$q_0$ is the
amplitude of the classical excitation of $q_1$
\begin{equation}
q_0 = - {U_0\over \Omega\rho}\;,
\end{equation}
and the constant
classical force $- (\rho\omega_e/2d) q_0^2$
[second term in Eq.\ (\ref{eomx})]
is applied to the test mass to counteract the mean radiation pressure
force [time average of first term in Eq.\ (\ref{eomx})].
\end{itemize}

One can take two points of view on the quantities $q_1$, $q_2$, $x$, and
$U_e$: one can regard them as classical quantities, with $U_e(t)$
described by a classical spectral density
$S_{U_e}(\omega)$, in which case
Eqs.\ (\ref{eom}) are classical equations of motion; or one can regard them as
quantum mechanical operators in the Heisenberg picture, in which case Eqs.\
(\ref{eom}) are the Heisenberg evolution equations.  The two viewpoints will
produce the same final conclusions, if one chooses the correct
quantum-mechanically-based value for $S_{U_e}$.  We shall return to
this in Sec.\ \ref{sec:spectraldensities} below.

We resolve $q_1$ and $q_2$ into their quadrature components

\begin{eqnarray}
q_1 &=& (q_0 + a_1)\cos\omega_e t + b_1 \sin \omega_e t\;, \nonumber\\
q_2 &=& a_2\cos\omega_e t + \left( -{\delta_e\over\Omega}q_0 +
b_2\right) \sin \omega_e t\;.
\label{q1q2}
\end{eqnarray}
Note that the classical input driving voltage $2U_0\cos\omega_e t$,
acting on resonator 2, produces its primary classical excitation
$q_0\cos\omega_e t$ in resonator 1 as was advertised in Sec.\
\ref{sec:intro}; but there is also a secondary classical excitation in
resonator 2 proportional to the loss rate $\delta_e$ that was ignored in
Sec.\ \ref{sec:intro}.

The quadrature amplitudes $a_{1,2}$ and $b_{1,2}$ carry
the perturbations caused by coupling to the test-mass position and to the
output waveguide.  We solve for these perturbations by inserting expressions
(\ref{q1q2}) into the equations of motion (\ref{eom}) and linearizing:

\begin{mathletters}
\label{no_losses_1}
\begin{eqnarray}
  \frac{d a_1(t)}{dt} + \delta_e a_1(t) & = &
    - \Omega b_2(t) - \frac{U_{es}(t)}{\rho}\;, \label{a1dot}\\
  \frac{d b_1(t)}{dt} + \delta_e b_1(t) & = &
    \frac{\omega_e q_0}{d}x(t) +
    \Omega a_2(t) + \frac{U_{ec}(t)}{\rho}\;,   \label{b1dot} \\
  \frac{d a_2(t)}{dt} & = & - \Omega b_1(t)\;,       \label{a2dot} \\
  \frac{d b_2(t)}{dt} & = & \Omega a_1(t)\;,  \label{b2dot} \\
  m{d^2 x\over dt^2} &=& F_{\rm BA}(t) + F_s(t)\;.
\label{xdotdot}
\end{eqnarray}
\end{mathletters}
Here $U_{ec}$ and $U_{es}$ are the quadrature amplitudes of the
fluctuating voltage imposed on resonator 1 from the output waveguide,
\begin{equation}
U_e = U_{ec} \cos\omega_e t + U_{es} \sin\omega_e t\;,
\label{UcUs}
\end{equation}
and $F_{\rm BA}(t)$ is the back-action force that the speed meter exerts on the
test mass
averaged over a microwave period $2\pi/\omega_e$,
\begin{equation}
F_{\rm BA}(t) = {q_0\rho\omega_e\over d} a_1(t).
\label{FBA}
\end{equation}
In the Heisenberg-picture interpretation of Eqs.\ (\ref{no_losses_1}),
all the functions of time $t$ are quantum operators except the classical
force $F_s(t)$.

We solve Eqs.\  (\ref{no_losses_1}) in the frequency domain using the
Fourier-transform conventions of Eq.\ (\ref{FTConvention}).
The frequencies of interest are in the range $|\omega|\ll\omega_e$
and can be thought of as side-band frequencies of the microwave carrier
$\omega_e$.
Equations (\ref{no_losses_1}) imply, for the quadrature amplitudes of
resonator 1:
\begin{eqnarray}
  a_1(\omega) &=& \frac{i\omega U_{es}(\omega)}{\rho{\cal L}(\omega)}
   \;,\nonumber \\
  b_1(\omega) &=& -\frac{i\omega}{{\cal L}(\omega)}\left(
    \frac{\omega_e q_0}{d}x(\omega)
    + \frac{U_{ec}(\omega)}{\rho}\right),
  \label{C1}
\end{eqnarray}
where
\begin{equation}
{\cal L}(\omega) \equiv \Omega^2-\omega^2-i\omega\delta_e\;.
\label{calL}
\end{equation}

The output-wave voltage entering the output waveguide can be expressed in
the form
\cite{yurke}:
\begin{eqnarray}
  U^{out}(t) &=& U_e(t) -
      \frac{2\delta_e}{\omega_e}\rho\frac{d q_1(t)}{dt} \nonumber \\
    &=& \left[U_{ec}(t) - 2\delta_e\rho b_1(t)\right]\cos\omega_e t
       \nonumber\\
  &&+ \left[U_{es}(t) + 2\delta_e\rho a_1(t)\right]\sin\omega_e t\;,
\label{C2}
\end{eqnarray}
where we have ignored the carrier signal $2\delta_e\rho q_0 \sin\omega_e t$.
When measuring the classical signal force $F_s(t)$, the noise will be
minimized by monitoring the sidebands of an optimally chosen quadrature
component of the output wave.  This monitoring can be done via homodyne
detection 
[which, at microwave frequencies, can be achieved by
mixing the output wave $U^{\rm out}(t)$ with a strong
local-oscillator field $U_{LO}\propto \sin(\omega_e t+\Phi)$,
where $\Phi$ is
the desired quadrature's phase, 
then rectifying it and averaging it over a carrier period, 
and then monitoring its slowly oscillating voltage].  
The monitored voltage is then proportional to
\begin{eqnarray}
  \tilde U(t) &=& \left[U_{ec}(t)
    - 2\delta_e\rho b_1(t)\right]\sin\Phi \nonumber\\
    &&+ \left[U_{es}(t) + 2\delta_e\rho a_1(t)\right]\cos\Phi\;.
\end{eqnarray}
By switching to the frequency domain and using expression
(\ref{C1}) for $b_1(\omega)$, we obtain the following expression for this
monitored voltage in terms of the test-mass position $x(\omega)$ and the
noise $x_{\rm m}(\omega)$ added to the position signal by the
speed meter:
\begin{equation}
  \tilde U(\omega) =
    \frac{2i\omega\omega_e\delta_e\rho q_0 \sin{\Phi}}{{\cal L}(\omega)d}
    \left(x(\omega)+x_{\rm m}(\omega)\right),
   \label{tildeU}
\end{equation}
where
\begin{eqnarray}
  x_{\rm m}(\omega) = \frac{d}{2i\omega\omega_e\delta_e q_0\rho}
    (\Omega^2-\omega^2+i\omega\delta_e) \nonumber\\
    \times \left[U_{ec}(\omega) +
    U_{es}(\omega)\cot\Phi\right]\;.
  \label{x_fl}
\end{eqnarray}

Notice that in the limit of weak coupling to the output waveguide
$\delta_e\ll\Omega$, and for signal frequencies low compared to the resonator
sloshing frequency $\omega\ll\Omega$, the monitored voltage is $\tilde U(t)
\propto [dx/dt + dx_{\rm m}/dt]$; i.e., it is proportional to the
test-mass velocity, as expected for a speed meter.  However, as we shall see
below, the regime of optimal sensitivity is one in which the classical
force's signal frequency is at $\omega \sim \Omega$, so the monitored voltage
(\ref{tildeU}) has a more complicated dependence on test-mass position than
simply $dx/dt$.

Equation (\ref{xdotdot}) implies for the test-mass position in the frequency
domain
\begin{equation}
x(\omega) = x_o \delta(\omega) - {p_o\over i\omega m}\delta(\omega)
- {F_{\rm BA}(\omega) \over m\omega^2} - {F_s(\omega) \over m\omega^2}\;.
\label{xomega}
\end{equation}
Here $x_o$ and $p_o$ are integration constants (the test-mass position
and momentum in the absence of coupling to the signal force $F_s(\omega)$
and to the speed meter), $\delta(\omega)$ is the Dirac delta function, and
\begin{equation}
F_{\rm BA}(\omega) =
    \frac{i\omega\omega_e q_0 U_{es}(\omega)}{{\cal L}(\omega)d}
  \label{F_fl}
\end{equation}
is the speed meter's back-action force; cf.\ Eqs.\ (\ref{FBA}) and (\ref{C1}).

\subsection{Meter and Back-Action Spectral Densities}
\label{sec:spectraldensities}

When thinking about this speed meter in the quantum mechanical Heisenberg
Picture, one might be concerned that the nonzero value of the test mass's
two-time commutator
$[x(t),x(t')] = i\hbar(t'-t)/m$ will cause the two-time
commutator of the output waveguide's signal
to be nonzero; cf.\ Eq.\
(\ref{tildeU}).  If this were so, then we would have to worry about the
effects of successive quantum state reductions as each successive bit of
signal is collected (via homodyne detection).  Fortunately, the monitored
quantity is the Hermitian part
$\tilde U^h(t) = {1\over2}\left(\tilde U(t) + \tilde U^{\dag}(t)\right)$ of
a quadrature amplitude $\tilde U(t)$ of the
output waveguide's microwave field $U^{\rm out}(t)$.
The commutation relations for the electromagnetic field guarantee
that this quantity commutes with itself at different
times $[\tilde U^h(t),\tilde U^h(t')] = 0$, independently of how the field
has interacted with the speed meter.
(This is a manifestation of the quantum Markov approximation.)  In the
case of the speed meter, this vanishing commutator is achieved via an automatic
cancellation between the influences of the test-mass position $x(t)$
[which in turn is influenced by the meter's back-action noise $F_{\rm BA}(t)$]
and the meter's
noise $x_{\rm m}(t)$; cf.\ Eqs.\ (\ref{tildeU})--(\ref{F_fl}).

Because $[\tilde U^h(t), \tilde U^h(t')] = 0$, we can compute
the noise in
any measurement with the speed meter by taking expectation values in the
initial states of the test mass, resonators, and incoming
output-waveguide field
$U_e$.  Moreover, when --- as in this paper --- we are not interested
in making absolute measurements of test-mass position and momentum, but
instead are interested only in learning about components of the classical
force $F_s(t)$ bounded away from zero frequency, our final inferred
force and its noise will be independent of the initial test-mass position and
momentum $x_o$ and $p_o$ [cf.\ Eq.\ (\ref{xomega}) where $x_o$ and $p_o$ appear
only at zero frequency].  In addition, in this section's model, which
ignores the resonators' intrinsic losses, the resonator dissipation via
leakage of field into the output waveguide guarantees that the state of the
resonators is determined completely by the initial state of the output
waveguide field $U_e$.

These considerations imply that the measurement noise will be determined solely
by the quantum state of the field $U_e$ that impinges on the speed
meter from the output waveguide.  Throughout this paper, except in Sec.\
\ref{sec:realization},
we shall assume that this field is in its vacuum state.
Correspondingly, the spectral densities and cross spectral density
of its quadrature components are
\begin{equation}
S_{U_c}(\omega) = S_{U_s}(\omega) = \hbar \rho\delta_e\;, \quad S_{U_c U_s} =
0\;.
\label{SUcs}
\end{equation}
(To deduce these spectral densities from the standard theory of a quantized
transmission line or waveguide, one must know that, in the notation of our
model, the waveguide impedance is $2 \rho \delta_e / \omega_e$.)

By combining Eqs.\ (\ref{SUcs}), (\ref{FBA}) and (\ref{x_fl}) we deduce
for the spectral densities of the meter's position noise and back-action force
and their cross spectral density
\begin{mathletters}
\label{sp_dens}
\begin{eqnarray}
  S_x(\omega) &=&
    \frac{\hbar|{\cal L}(\omega)|^2}{2m\omega^2\Lambda^4\sin^2\Phi}\;, \\
  S_F(\omega) &=&
    \frac{\hbar m\Lambda^4\omega^2}{2|{\cal L}(\omega)|^2}\;, \\
  S_{x F}(\omega) &=& -\frac{\hbar}{2}\cot\Phi\;.
\end{eqnarray}
\end{mathletters}
Here $\Lambda$ is a frequency that characterizes the strength of the pumping,
\begin{equation}
\Lambda^4 \equiv \frac{2\omega_e W}{md^2}
\label{Lambda}
\end{equation}
with
\begin{equation}
W = \rho \omega_e q_0^2 \delta_e
\end{equation}
the power supplied to the resonator by the input waveguide and the
corresponding power removed through the output waveguide; cf.\ Eq.\
(\ref{Ej}).  Below it will be useful to write $|{\cal L}(\omega)|^2$ [Eq.\
(\ref{calL})] in the form
\begin{equation}
  |{\cal L}(\omega)|^2 = (\omega^2-\omega_0^2)^2 +
\delta_e^2(\omega_0^2 + \delta_e^2/4)\;,
\label{abscalL}
\end{equation}
where
\begin{equation}
\omega_0 \equiv \sqrt{\Omega^2-\delta_e^2/2}\;;
\label{omegaO}
\end{equation}
$\omega_0$ will turn out to be the speed meter's optimal frequency of
operation.

\subsection{Wide-band sensitivity with lossless resonators}
\label{sec:WB_lossless_sensitivity}

When one infers the classical signal force $F_s(t)$ from the speedmeter's
output $\tilde U(t)$,
the spectral density of the noise of the inferred $F_s$ is
\begin{eqnarray}
  S(\omega) &=&
    m^2\omega^4 S_x(\omega) -2m\omega^2 S_{xF}(\omega) +
    S_F(\omega) \nonumber \\
  &=& \hbar m\omega^2\xi_{\rm WB}^2(\omega);
  \label{S_total}
\end{eqnarray}
cf.\ Eq.\ (\ref{xiWBdef}).
Equations (\ref{sp_dens})
and (\ref{S_total})
imply for the amount by which the speed meter
beats the wide-band standard quantum limit
\begin{equation}
  \xi_{\rm WB}^2(\omega) = \frac{|{\cal L}(\omega)|^2}{2\Lambda^4\sin^2\Phi} +
    \cot\Phi + \frac{\Lambda^4}{2|{\cal L}(\omega)|^2} .
\end{equation}

We shall optimize the homodyne phase $\Phi$ so as to minimize $\xi_{\rm WB}$ at the
frequency $\omega_F$ around which the signal force $F_s(t)$ is concentrated.
The optimizing phase is
\begin{equation}
  \cot\Phi = -\frac{\Lambda^4}{|{\cal L}(\omega_F)|^2}\;,
\end{equation}
$\xi^2_{\rm WB}(\omega)$ for this $\Phi$ is
\begin{equation}
  \xi_{\rm WB}^2(\omega) = \frac{|{\cal L}(\omega)|^2}{2\Lambda^4} +
    \frac{\Lambda^4(\omega^2-\omega_F^2)^2(\omega^2+\omega_F^2-2\omega_0^2)^2}
      {2|{\cal L}(\omega)|^2|{\cal L}(\omega_F)|^4} \;,
\end{equation}
and its minimum is
\begin{equation}
  \xi^2_{\rm min} = \xi_{\rm WB}^2(\omega_F) =
    \frac{(\omega_F^2-\omega_0^2)^2+\delta_e^2(\omega_0^2+\delta_e^2/4)}
      {2\Lambda^4}\;.
\end{equation}
To further minimize the noise, we shall adjust the speed meter's
optimal frequency to $\omega_0 = \omega_F$, thereby producing
\begin{equation}
  \xi_{\rm WB}^2(\omega) = \frac{|{\cal L}(\omega)|^2}{2\Lambda^4} +
    \frac{\Lambda^4(\omega^2-\omega_0^2)^4}
      {2|{\cal L}(\omega)|^2\delta_e^4(\omega_0^2+\delta_e^2/4)^2}\; ,
  \label{xi_omega}
\end{equation}
and
\begin{equation}
  \xi^2_{\rm min}  = \frac{\delta_e^2(\omega_0^2+\delta_e^2/4)}{2\Lambda^4}
   = {W_{\rm SQL}\over W}\;,
  \label{xi_min}
\end{equation}
where
\begin{equation}
W_{\rm SQL} = {md^2(\omega_0^2 + \delta_e^2/4)\delta_e^2 \over 4\omega_e}
\label{WSQL}
\end{equation}
is the pump power required to reach the standard quantum limit at the
optimal frequency $\omega_0$.  By pumping with a power $W>W_{\rm SQL}$, the
speed meter can beat the SQL in the vicinity of the optimal frequency
$\omega_0$.

We define the frequency band $\omega_1 < \omega < \omega_2$ of high
sensitivity to be those frequencies for which
\begin{equation}
  \xi_{\rm WB}(\omega) \le \sqrt2\xi_{\rm WB}(\omega_0). \label{range}
\end{equation}
From Eqs.\ (\ref{xi_omega}) and ({\ref{xi_min}), we infer that
\begin{equation}
  \omega_{1,2}^2 = \omega_0^2 \mp
    \frac{\delta_e^2(\omega_0^2+\delta_e^2/4)}
      {\sqrt[4]{\delta_e^4(\omega_0^2+\delta_e^2/4)^2+\Lambda^8}} =
    \omega_0^2 \mp \frac{2\Lambda^2\xi_{\rm min}^2}{\sqrt[4]{4\xi_{\rm min}^4+1}}
\label{omega12}
\end{equation}
Equations (\ref{omega12}),
(\ref{xi_min}) and
(\ref{xi_omega}) imply that the lossless
speed meter can
beat the Force-measurement SQL by a large amount $\xi_{\rm min}\ll 1$ over a wide
frequency band, $\omega_2 - \omega_1 \sim \omega_F$ by setting
$\Lambda/\omega_0 \sim (\delta_e/\omega_0)^2 \agt 2$; cf. Fig.\ \ref{fig:xiWB}
in Appendix \ref{app:comparison}.

\subsection{Narrow-band sensitivity with lossless resonators}
\label{sec:NB_lossless_sensitivity}

At fixed pump power $W$, i.e.\ fixed $\Lambda$, Eqs.\
(\ref{omega12}) and (\ref{xi_omega}) imply that there is a trade off,
as one changes $\delta_e$, between
the optimal sensitivity $\xi_{\rm min}$ and the frequency band $\omega_2 -
\omega_1$ of near-optimal sensitivity.  For $\delta_e \to
0$ the sensitivity at $\omega_0$ grows indefinitely, but the frequency band
goes to zero. If
$\xi_{\rm min} \ll 1$ and $|\omega_2-\omega_1| \ll \omega_0$, this tradeoff has
a simple form:
\begin{equation}
  \frac{\omega_2-\omega_1}{\omega_0} =
    2\left(\frac{\Lambda}{\omega_0}\right)^2\xi_{\rm min}^2 =
  \sqrt{\frac{8\omega_e W}{md^2\omega_0^4}}\xi_{\rm min}^2 .
\end{equation}

In this narrow-band regime (more precisely, for $\delta_e \ll \omega_0$ and for
a frequency range $\Delta\omega \ll \delta_e^2/\omega_0$ centered on
$\omega_0$), the spectral density of the net
noise has the form [Eqs.\ (\ref{S_total}) and (\ref{xi_omega})]
\begin{equation}
S(\omega) = A'(\omega_{\rm meter}^2-\omega^2)^4 + B'\;,
\label{SAprimeBprime}
\end{equation}
where $\omega_{\rm meter} = \omega_0$ and
\begin{equation}
A' = {\hbar m \Lambda^4\over 2\omega_0^4\delta_e^6}\;, \quad
B'={\hbar m \omega_0^4\delta_e^2\over2\Lambda^4}\;.
\label{AprimeBprime}
\end{equation}
Notice that for the narrow-band speed meter, the noise's frequency dependence
[Eq.\ (\ref{SAprimeBprime})] is $(\omega_{\rm meter}^2-\omega^2)^4$, whereas for an
ordinary, quantum limited meter [Eq.\ (\ref{narrow_band_S})] it is
$(\omega_{\rm meter}^2-\omega^2)^2$.  The $(\omega_{\rm meter}^2-\omega^2)^4$
behavior is responsible for the ability of the speed meter to beat the
narrow-band SQL, and is produced by the combined actions of the speed meter's
multiple degrees of freedom (test mass and two resonators) and the correlation
$S_{xF}\ne 0$ of its noises.  These combined actions make the net noise
$S(\omega) = m^2 \omega^4 \xi_{\rm WB}^2(\omega)$
be equivalent to\footnote{For a
detailed discussion of the use of noise correlations to make a meter's noise
resemble that of a system that has different dynamical motions than the
meter actually possesses, see Ref.\ \cite{syrtsev_khalili}. Section
\ref{sec:narrowbandSQL} above gives another example: the noise correlation
is used there to make the noise be that of an oscillator with eigenfrequency
$\omega_{\rm meter}$ different from the oscillator's true frequency $\omega_0$.}
that of a system which
has two coupled dynamical degrees of freedom with system eigenfrequencies
that are degenerate [equation of motion of the quartic form $d^4 y/dt^4 +
2 \omega_{\rm meter}^2
d^2 y/dt^2 + \omega_{\rm meter}^4 y = F_s(t)$ for some variable $y$.]
The noise-equivalence to such a system is the
central feature of a measuring device that beats the narrow-band SQL.  (Of
course, one can do even better with a device whose noise behaves like
that of a system with three degenerate eigenfrequencies.)

Three of the authors have previously described a conceptual design
for an ``optical-bar'' gravitational-wave antenna \cite{optbar} that
can beat the gravitational-wave narrow-band SQL and does so by this same
principle, but without the aid of noise correlations.
When operating in a narrow-band mode, the optical bar
does actually consist of two coupled degrees of freedom with
system eigenfrequencies
that are degenerate, and it thus does actually have
the above, quartic equation of
motion.\footnote{The degrees of freedom are (i) the electromagnetic energy
that sloshes between the two nearly identical Fabry-Perot
cavities (energy difference $\delta {\cal E}(t)$, and (ii) the
displacement $y(t) = x_D - (x_A + x_B)/2$ of
the cavities' common corner mirror $D$ relative to the separate end mirrors
$A$ and $B$; see Fig.\ 1a of Ref.\ \protect\cite{optbar}.}

For the speed meter, Eqs.(\ref{AprimeBprime}) imply that
\begin{equation}
A'B' = {\hbar^2 m^2\over4\delta_e^4}\;.
\label{AprimeTimesBprime}
\end{equation}
This relation, together with Eq.\ (\ref{SAprimeBprime}),
implies that, when a measurement of a sinusoidal force with 
$\omega_F = \omega_{\rm meter}$ and duration $\tau_F$ is made by 
averaging over a time $\hat\tau \agt \tau_F$, and the ratio $B'/A'$ is
optimized to $B'/A'\sim (\omega_F/\hat\tau)^4$,
then the amplitude of the minimum detectable classical force is
\begin{equation}
F \simeq {\sqrt{\hbar m \omega_F}\over \tau_F} \sqrt{\omega_F/\delta_e \over
\delta_e \hat\tau} = F_{\rm NB\;SQL} \sqrt{\omega_F/\delta_e \over
\delta_e \hat\tau}\;,
\end{equation}
which beats the narrow-band SQL (\ref{NB_SQL}) by the indicated factor.
This result can also be derived by comparing Eqs.\ (\ref{narrow_band_S}),
(\ref{AB}), (\ref{SAprimeBprime}) and
(\ref{AprimeBprime}) to obtain for the amount by which the
narrow-band SQL is beaten at frequency $\omega$
\begin{equation}
\xi_{\rm NB}^2=\frac{\omega^2_F-\omega^2}{\delta^2_e},
\end{equation}
and by then evaluating the rms value of $\xi_{\rm NB}$ over the bandwidth
$\Delta\omega = 2\pi/\hat\tau$ of the measurement to obtain
\begin{equation}
\xi_{\rm NB}^{\rm rms} \simeq \sqrt{\omega_F/\delta_e \over
\delta_e \hat\tau}\;.
\end{equation}

\section{The sensitivity of the speedmeter with intrinsic losses}
\label{sec:losses}

Turn, now, from the idealized case of a speed meter with no intrinsic
losses in its resonators to the more realistic case of lossy resonators.
In this case the resonators' equations of motion become
\begin{mathletters}
  \label{with_losses}
\begin{eqnarray}
  &&\frac{d^2 q_1(t)}{dt^2} + 2(\delta_1+\delta_e)\frac{d q_1(t)}{dt} +
    \omega_e^2\left(1-\frac{x(t)}{d}\right)^2 q_1(t) \nonumber \\
  &&\quad\quad\quad =
    2\omega_e\Omega q_2(t) + 2 \frac{\omega_e}{\rho}\left[ U_1(t) +
    U_e(t)\right]\;,
     \\
  &&\frac{d^2 q_2(t)}{dt^2} + 2\delta_2\frac{d q_2(t)}{dt} +
    \omega_e^2q_2(t) \nonumber \\
  &&\quad\quad\quad =
    2\omega_e\Omega q_1(t) + 2\frac{\omega_e}{\rho}U_0\cos\omega_e t +
    2\frac{\omega_e}{\rho}U_2(t) ,
\end{eqnarray}
\end{mathletters}
where $\delta_{1,2}$ are the rates of amplitude decay in resonators
1 and 2 due to intrinsic losses and $U_{1,2}$ are the fluctuating
voltages that must accompany these losses.

Inserting expressions (\ref{q1q2}) into these equations of motion and
linearizing, we obtain the following generalization of
Eqs.\ (\ref{no_losses_1}):
\begin{eqnarray}
  \frac{d a_1(t)}{dt} + (\delta_1+\delta_e) a_1(t) & = &
    - \Omega b_2(t) - \frac{U_{1s}(t)}{\rho}
    - \frac{U_{es}(t)}{\rho}\;,
    \nonumber \\
  \frac{d b_1(t)}{dt} + (\delta_1+\delta_e) b_1(t) & = &
    \frac{\omega_e q_0}{d}x(t) +
    \Omega a_2(t)\;, \nonumber \\
    && + \frac{U_{1c}(t)}{\rho} + \frac{U_{ec}(t)}{\rho}\;,
    \nonumber \\
  \frac{d a_2(t)}{dt} + \delta_2 a_2(t) & = &
    - \Omega b_1(t) - \frac{U_{2s}(t)}{\rho}\;. \nonumber \\
  \frac{d b_2(t)}{dt} + \delta_2 b_2(t) & = &
    \Omega a_1(t) + \frac{U_{2c}(t)}{\rho}
\label{with_losses1}
\end{eqnarray}
By repeating the same manipulations as in section \ref{sec:no_losses}
and using
\begin{eqnarray}
&&S_{U_{js}}=S_{U_{jc}} = \hbar\rho\delta_j\;, \quad
S_{U_{js}U_{jc}} = 0\;, \nonumber \\
&&S_{U_{js}U_{ks}} = S_{U_{js}U_{kc}} =S_{U_{jc}U_{kc}} =0
\end{eqnarray}
for $j\ne k$ and  $j,k=1,2$ [cf.\ Eq.\ (\ref{SUcs})],
we obtain the following expressions
for the spectral densities of the speed meter's position noise and back-action
noise:
\begin{eqnarray}
  S_x(\omega) & = & \frac{\hbar|{\cal L'}(\omega)|^2}
    {2m(\omega^2+\delta_2^2)\Lambda^4\sin^2\Phi}
  \nonumber \\ \nonumber \\
  S_F(\omega) & = &
    \frac{\hbar m\Lambda^4
        \left[(\omega^2+\delta^2_2)(\delta_1+\delta_e)+\Omega^2\delta_2\right]}
      {2|{\cal L'}(\omega)|^2\delta_e}
  \nonumber \\ \nonumber \\
  S_{x F}(\omega) & = & -\frac{\hbar}{2}\cot\Phi,
  \label{sp_dens_losses}
\end{eqnarray}
where
\begin{equation}
  |{\cal L'}(\omega)|^2 =
    (\omega^2-{\omega_0'}^2)^2 + {\delta^*}^2
( {\omega_0'}^2 + {\delta^*}^2 /4)\;.
\end{equation}
Here $\delta^*$ is the total damping rate due to intrinsic losses and losses
into the output waveguide and $\omega_0'$ is the speed meter's
damping-influenced optimal frequency of operation
\begin{equation}
\delta^* = \delta_e + \delta_1 + \delta_2\;, \quad
  \omega_0' = \sqrt{\Omega^2-[(\delta_1+\delta_e)^2+\delta_2^2]/2}\;.
\label{omega0prime}
\end{equation}

By inserting the speed-meter spectral densities (\ref{sp_dens_losses})
into Eq.\ (\ref{S_total})), we obtain for the factor by which the
lossy speed meter can beat the classical-force standard quantum limit
\begin{eqnarray}
  \xi_{\rm WB}^2(\omega) &=&
    \frac{|{\cal L'}(\omega)|^2\omega^2}
      {2\Lambda^4\sin^2\Phi(\omega^2+\delta_2^2)} +
    \cot\Phi \nonumber\\
  &&+
    \frac{(\omega^2+\delta^2_2)(\delta_1+\delta_e)+\Omega^2\delta_2}
      {2\omega^2\delta_e|{\cal L'}(\omega)|^2}
    \Lambda^4\; .
   \label{xi_WB_lossy}
\end{eqnarray}
To minimize the noise at the frequency $\omega_F$ around which the signal force
$F_s(t)$ is concentrated, we adjust
the speedmeter so $\omega_0' = \omega_F$ and choose for the homodyne phase
\begin{equation}
  \cot\Phi = -\frac{\Lambda^4}{|{\cal L'}(\omega_0')|^2}
    \frac{{\omega_0'}^2+\delta_2^2}{{\omega_0'}^2}\;.
\end{equation}
The result is
\begin{eqnarray}
  \xi_{\rm WB}^2(\omega_0') &=&
    \frac{\delta^{*2}({\omega_0'}^2+\delta^{*2}/4){\omega_0'}^2}
      {2\Lambda^4({\omega_0'}^2+\delta_2^2)} \nonumber \\
  &&+
    \frac{\Lambda^4[\delta_1({\omega_0'}^2+\delta_2^2)+\Omega^2\delta_2]}
      {2{\omega_0'}^2\delta_e\delta^{*2}
        ({\omega_0'}^2+\delta^{*2}/4)} \;.
\label{xiWB_lossy}
\end{eqnarray}
By contrast with the lossless case, the sensitivity here
does not grow indefinitely with the growth of $\Lambda$.  Rather, the
sensitivity at the optimal frequency $\omega_0'$ is maximized by setting
\begin{equation}
  \Lambda^4 =
    \frac{{\omega_0'}^2\delta^{*2}
        ({\omega_0'}^2+\delta^{*2}/4)\sqrt\delta_e}
      {\sqrt{\left[({\omega_0'}^2+\delta_2^2)\delta_1+\Omega^2\delta_2\right]
        ({\omega_0'}^2+\delta_2^2)}} \;.
\end{equation}
In this case
\begin{equation}
  \xi_{\rm min}^2 = \xi_{\rm WB}^2({\omega_0'}) = \sqrt{\frac{\delta_1}{\delta_e} +
    \frac{{\omega_0'}^2+{1\over2}\left[(\delta_1+\delta_e)^2+\delta_2^2\right]}
    {{\omega_0'}^2+\delta_2^2}
      \frac{\delta_2}{\delta_e}}
\label{xires}
\end{equation}
In any real speed meter, one will make the losses $\delta_1$ and $\delta_2$ as
small as one can, resulting in
$\delta_1\simeq \delta_2\ll \delta_e, \omega_0'$.  This further simplifies
expression
(\ref{xires}) into the form
\begin{equation}
  \xi_{\rm min}^4 = \frac{2\delta_1}{\delta_e} +
    \frac{\delta_e \delta_1}{2{\omega_0'}^2}\;,
\label{xires2}
\end{equation}
which is optimized by setting $\delta_e=2\omega_0'$ [so $\Omega = \sqrt3
\omega_0'$; cf.\ Eq.\ (\ref{omega0prime})]:
\begin{equation}
  \xi_{\rm min}^4 = \frac{2\delta_1}{\omega_0'} = {4\delta_1\over \delta_e}\;.
\label{xires3}
\end{equation}
In this case the actual (optimal) pump power $W$ and power to reach the
SQL, $W_{\rm SQL}$, are
\begin{equation}
W = \frac{\Lambda^4 m d^2}{2\omega_e} = {W_{\rm SQL}\over \xi_{\rm min}^2}\;, \quad
W_{\rm SQL} = \frac{4md^2{\omega_0'}^4}{\omega_e}\;,
\label{power}
\end{equation}
the homodyne phase is $\cot\Phi = -1/\xi_{\rm min}^2$, and the band over which
$\xi_{\rm WB}^2 < 2 \xi_{\rm min}^2$ is
\begin{equation}
{\omega_2 - \omega_1 \over \omega_0'} = 2\sqrt[4]{8} \xi_{\rm min}\;.
\end{equation}
Of course, 
by allowing the minimum of $\xi_{\rm WB}^4(\omega)$ to be larger than
$4\delta_1/\delta_e$, one can widen the band of good sensitivity to
$\omega_1 - \omega_2 \sim \omega_0'$, as in the case of the lossless
speed meter [Eq.\ (\ref{omega12}) and associated discussion; Fig.\
\ref{fig:xiWB} of Appendix \ref{app:comparison}].

\section{On the possibility to realize the quantum speedmeter}
\label{sec:realization}

We turn, now, to a discussion of the possibility to construct a
demonstration version of the quantum speed meter that is capable of beating
the wide-band SQL.

A central issue in such a speed meter is the intrinisic losses
in the resonators.  These losses are characterized by the dissipation rate
$\delta_1 \simeq \delta_2$, or equivalently by the unloaded resonators'
energy damping time $\tau_1 = 1/(2\delta_1)$ or
quality factor $Q_1 = \omega_e \tau_1$.
Equations (\ref{xires3}) and (\ref{power}) show that the
intrinsic damping time $\tau_1$ can seriously
limit the achievable $\xi_{\rm min} = 1/(\omega_0' \tau_1)^{1/4}$ and
significantly influence the required pump power
$W = W_{\rm SQL}\sqrt{\omega_0' \tau_1}$ and the power that is thermally
dissipated in each resonator, $W' = (\delta_1/\delta_e) W =
W/(4\omega_0'\tau_1) = W_{\rm SQL}/(4\sqrt{\omega_0'\tau_1})$.

Actually, the situation is more extreme than these equations suggest.
Even at cryogenic temperatures 
$T\simeq 1K$,
the mean thermal energy per degree of freedom
$kT$ is large compared to the energy of a microwave photon $\hbar\omega_e$;
i.e., the thermal noise number
\begin{equation}
N_T = {kT\over \hbar\omega_e} \simeq 2
\end{equation}
is somewhat larger than unity.  
(Here and below, for reasons to be discussed,
we set $\omega_e = 2\pi \times 10^{10} {\rm s}^{-1}$.) Correspondingly,
the quantum-to-classical transition $\hbar\omega_e/2 \rightarrow kT$
implies that the
noise spectra of the fluctuating voltages $S_{U_e}$, $S_{U_1}$ and $S_{U_2}$
that plague
the speed meter are larger by $2N_T$ than in the idealized, quantum-limited
analysis of Secs.\ \ref{sec:no_losses} and \ref{sec:losses}, and
the limiting performance and thermally dissipated power are changed by
factors $\sqrt{2N_T}$ and $2N_T$:
\begin{equation}
\xi_{\rm min} = {\sqrt{2N_T}\over(\omega_0'\tau_1)^{1/4}}\;, \quad
W' = {2N_T\, md^2\omega_0'^4 \over \omega_e \sqrt{\omega_0' \tau_1}}\;.
\label{necessary}
\end{equation}
(Here we have used Eq.\ (\ref{power}) for $W_{\rm SQL}$.)

To have any hope at all of achieving $\xi_{\rm min} <1$, it is necessary to operate
at cryogenic temperatures 
$T\simeq 1{\rm K}$.  
Then, for a demonstration
experiment that achieves $\xi_{\rm min} \simeq 0.5$ near a frequency
$\omega_0' \simeq 3\times 10^3 {\rm s}^{-1}$
for the signal force, Eq.\ (\ref{necessary})
dictates a resonator energy damping time $\tau_1 \simeq 0.1 {\rm s}$,
corresponding to an unloaded quality factor
$Q_1 \simeq 5\times 10^{9}$.

The best candidates for resonators with
$Q_1 \simeq 5\times 10^{9}$
are polished sapphire
disks excited in whispering-gallery modes with $\omega_e \sim
2\pi\times 10^{10} {\rm s}^{-1}$ (which is our reason for selecting this
$\omega_e$). Such resonators have
been constructed with $Q_1$ larger than
$10^9$ \cite{Whisp}, and the intrinsic electromagnetic losses in sapphire
are small enough to permit $Q_1 \simeq 10^{10}$.  Moreover,
the whispering-gallery evanescent fields provide an attractive
means for coupling to the test mass and to input and output waveguides.
To obtain a small tuning length $d$, resonator 1 and the test mass could
consist of two identical disks $A$ and $B$ facing each other with variable
separation [and $x = ($change of separation)], with the resonator-1
whispering-gallery
field shared equally between the disks, and with the classical force $F_s(t)$ acting on
$A$; while resonator 2 could be a single disk $C$ facing $B$ and with fixed
separation from $B$ large enough that the fields in $C$ and in $AB$ overlap
only slightly.  In this case, the tuning length $d$ can be as small as
$d \simeq 3 {\rm mm}$ \cite{MW_resonator,cuthbertson} but not smaller.
So large a $d$ means that each resonator's
thermally dissipated power (\ref{necessary}) will be,
for $m = 10 {\rm g}$ (the smallest reasonable test mass corresponding to
the smallest dissipated power) and all other parameters as above,
$W'\sim 3 \times 10^2 {\rm erg}/{\rm s}$.
So much heat cannot be removed
radiatively, but it can be removed by thermal conduction up the suspensions
from which the test mass and resonators hang, provided the suspensions are
thin niobium strips rather than the more normal fused-silica fibers.

To achieve a demonstration experiment with $\xi_{\rm min} \simeq 0.5$, the
test mass's thermal mechanical noise must be kept correspondingly small:
\begin{equation}
  \frac{2kT}{\tau^*_m} < \xi_{\rm min}\,\hbar\omega_0'\;,
\end{equation}
where $\tau^*_m$ is the test mass's mechanical relaxation time.  For the
above parameters, this will be satisfied if
$\tau^*_m > 2\times 10^8 {\rm s}^{-1}$.
Mitrofanov and colleagues \cite{Mitrofanov} have demonstrated
$\hat\tau^*_m$ comparable to this with
fused silica suspensions, and a similar performance is likely from a niobium
strip suspension \cite{Blair}.

The demonstration experiment also requires that
the meter measure the test-mass velocity
\begin{equation}
  \Delta v = \xi_{\rm min}\sqrt{\frac{\hbar}{m\hat\tau}}\;,
\end{equation}
where $\hat\tau \sim 1/\omega_0'$ is the observation time.  The above
parameters give
$\Delta v\sqrt{\hat\tau}\simeq 5\times 10^{-15}{\rm cm/s}^{1/2}$, a
signal strength that is within the measurement capabilities of current
techniques based on whispering gallery modes of sapphire disks and
microwave oscillators stabilized by sapphire disks\cite{cuthbertson}.

The velocity signal $\Delta v\sqrt{\hat\tau}\simeq 5\times 10^{-15}{\rm
cm/s}^{1/2}$ produces a microwave phase shift with magnitude
\begin{equation}
  \Delta\phi =  \frac{\omega_e}{\sqrt8 \,\omega_0'^2 d}\Delta v\;;
\end{equation}
i.e., for the above parameters,
$\Delta\phi\sqrt{\hat\tau} = 4\times 10^{-11}{\rm s}^{1/2}$.
This small phase shift imposes very strict requirements on the
stability of the microwave oscillator that regulates the speed meter's pump
field, though the quantum limit in this
case is not the main factor.  That stability translates into an oscillator
power
\begin{equation}
  W_{\rm osc} > \frac{8md^2{\omega_0'}^2\omega_e}{\xi_{\rm min}^2Q^2},
\end{equation}
where $Q$ is the quality factor of its resonator. For $Q=10^9$, the required
power is
$W_{\rm osc} >20 {\rm erg/s}$, which is within current technical
capabilities.

Thermal noise in the acoustic modes of the speed meter's resonators must
also be taken into account.
During the observation time $\hat\tau$, the thermally
induced change in the velocity that is measured by the speed meter will be
\begin{equation}
  \Delta v_{\rm ac} \simeq
    \omega_0'\sqrt{\frac {2 kT}{m\omega_{\rm ac}^3Q_{\rm ac}\hat\tau}}\;,
\end{equation}
where $\omega_{\rm ac}$ is the eigenfrequency and $Q_{\rm ac}$
the quality factor of
the lowest acoustic mode.
With the conservative estimate $Q_{\rm ac}=10^5$ at $\omega_{\rm ac}=10^6$, we
infer
$\Delta v_{\rm ac}\sqrt{\hat\tau}\simeq 5\times
10^{-17}{\rm cm/s}^{1/2}$,
which is small compared to the signal $\Delta v\sqrt{\hat\tau}\simeq 5\times
10^{-15}{\rm cm/s}^{1/2}$.

In summary, the above estimates suggest that
with present technology a demonstration type of
experiment at the level $\xi_{\rm min}\simeq 0.5$
is not hopeless.  However, further technological developments will be
required if such a speed meter is to become a promising tool for, e.g.,
QND interferometers in LIGO of the type proposed in Refs.\
\cite{optbar,symph}.  Most importantly, it will be necessary to construct
resonators with $Q_1 > 10^{10}$.  This may be possible for sapphire in double
disks (the design described above), or perhaps for klystron-type
superconducting resonators with lumped capacitances that permit tuning lengths
$d\sim 10^{-3} {\rm cm}$ (much smaller than the $d\simeq 3 {\rm mm}$
of sapphire disks).

\section*{Acknowledgments}
For helpful advice, KST thanks Andrey Matsko, Sergey Vyatchanin, and
the members of the Caltech QND Reading Group,
most especially Constantin Brif, Bill Kells,
Jeff Kimble, Yuri Levin and John Preskill.  This paper was
supported in part by NSF grants PHY--9424337, PHY--9503642 and PHY--9900776,
and by the Russian Foundation for Fundamental Research grants
\#96-02-16319a and \#97-02-0421g.

\appendix

\section{Comparison of Speed Meter and Position Meter}
\label{app:comparison}

It is useful to compare our speed meter (Fig.\ \ref{fig:speedmeter}) with
a position meter (parametric transducer) that is made from a single
microwave resonator,
modulated by the position of a test mass on which a signal force
acts; see Fig.\ \ref{fig:positionmeter}.

\begin{figure}
\unitlength=2pt

\begin{picture}(180,48)


\thinlines       
\put(15,36){\line(1,0){30}}\put(45,34){\oval(4,4)[r]}
\put(25,36){\vector(1,0){10}}
\put(30,34){\makebox(0,0)[ct]{$U_0\sin\omega_e t$}}

\thicklines      
\put(75,10){\line(0,1){35}}
\put(45,10){\line(0,1){35}}\put(45,45){\line(1,0){30}}
\put(80,45){\makebox(0,0)[cc]{\large{1}}}

\thicklines      
\put(46,10){\line(1,0){28}}\put(74,10){\line(0,1){5}}
\put(46,10){\line(0,1){5}}\put(46,15){\line(1,0){28}}
\put(46,10){\makebox(28,5)[cc]{\large{m}}}

\thinlines
\put(50,30){\vector(0,-1){15}}\put(50,30){\vector(0,1){15}}
\put(52,30){\makebox(28,5)[lc]{\large{d}}}

\thinlines       
\put(75,32){\line(1,0){35}}\put(75,30){\oval(4,4)[l]}
\put(77,36){\vector(1,0){10}}
\put(90,36){\makebox(0,0)[lc]{$U^{out}(t)$}}
\put(87,28){\vector(-1,0){10}}
\put(89,28){\makebox(0,0)[lc]{$U_e(t)$}}

\end{picture}

\caption{Schematic diagram of a position meter (parametric transducer).}
\label{fig:positionmeter}
\end{figure}
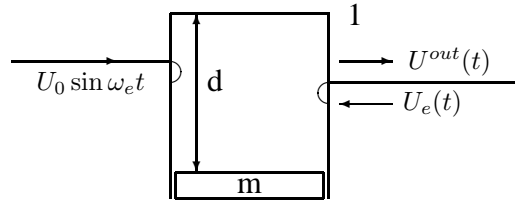

\subsection{Analysis of position meter}

The
position meter's
resonator is pumped with a classical force $U_0 \sin(\omega_e t)$,
by contrast with $U_0\cos(\omega_e t)$ for the speed meter; this difference
guarantees that the excitation in the resonator will be at the same phase
as for the speed meter's resonator $1$; see below.  The equations of motion for
the position meter are then the same as for the speed meter [Eqs.\
(\ref{with_losses})]
but with the driving voltage moved from resonator 2 to resonator 1 and changed
in phase so $\cos \to \sin$, with resonator 2 removed, and with the coupling
frequency $\Omega$ set to zero:
\begin{mathletters}
\label{eompos}
\begin{eqnarray}
  &&\frac{d^2 q(t)}{dt^2} + 2(\delta_e+\delta_1)\frac{d q(t)}{dt} +
    \omega_e^2\left(1-\frac{x(t)}{d}\right)^2 q(t) \nonumber\\
  &&\quad = 2\frac{\omega_e}{\rho}[U_1(t) + U_e(t) + U_0 \sin\omega_e t]\;,
      \label{eomqpos}\\
  &&m{d^2 x\over dt^2} = {\rho\omega_e\over d}q^2
     - {\rho\omega_e\over 2d}q_0^2 + F_s(t)\;.
      \label{eomxpos}
\end{eqnarray}
\end{mathletters}

Resolving $q_1$, $U_1$, and $U_e$ into $\cos\omega_e t$ and $\sin \omega_e t$
parts as for the speed meter [Eqs.\ (\ref{q1q2}), (\ref{UcUs}), etc.] and
linearizing, we obtain the same equations as for the speed meter [Eqs.\
(\ref{with_losses1})] but with resonator 2 deleted and $\Omega$ set to zero:
\begin{eqnarray}
  \frac{d a_1(t)}{dt} + (\delta_1+\delta_e) a_1(t) & = &
    - \frac{U_{1s}(t)}{\rho}
    - \frac{U_{es}(t)}{\rho}\;,
    \nonumber \\
  \frac{d b_1(t)}{dt} + (\delta_1+\delta_e) b_1(t) & = &
    \frac{\omega_e q_0}{d}x(t) 
     + \frac{U_{1c}(t)}{\rho} + \frac{U_{ec}(t)}{\rho}\;,
  \nonumber \\
\label{pos_with_losses1}
\end{eqnarray}

Repeating the same manipulations as for the speed meter, we arrive at
spectral densities for the position meter's position noise $x_{\rm m}(t)$ and
back-action noise $F_{\rm BA}(t)$, which can be deduced from those
(\ref{sp_dens_losses})
for the speed meter by setting $\Omega = 0$, $\delta_2 = 0$, and
therefore $|{\cal L}'(\omega)|^2 = \omega^2(\omega^2+{\delta^*}^2)$:
\begin{eqnarray}
  S_x(\omega) &=& \frac{\hbar(\omega^2+{\delta^*}^2)}
    {2m\Lambda^4\sin^2\Phi}\;, \quad
  S_F(\omega) =
    \frac{\hbar m\Lambda^4 \delta^*}
     {2(\omega^2+{\delta^*}^2)\delta_e}
  \nonumber \\ \nonumber \\
  S_{x F}(\omega) &=& -\frac{\hbar}{2}\cot\Phi,
  \label{sp_dens_losses_pos}
\end{eqnarray}
where $\delta^* = \delta_e + \delta_1$.  Correspondingly, when homodyne
detection is performed on the output of the position meter, with homodyne
angle $\Phi$, the factor by which
the the wide-band SQL is beaten is
[Eq.\ (\ref{xi_WB_lossy})]
\begin{equation}
  \xi_{\rm WB}^2(\omega) =
    \frac{\omega^2 (\omega^2+{\delta^*}^2)}
      {2\Lambda^4 \sin^2\Phi} +
    \cot\Phi + \frac{\delta^* \Lambda^4}
      {2\omega^2(\omega^2 + {\delta^*}^2)\delta_e}\;.
   \label{xi_WB_lossy_pos}
\end{equation}

\subsection{Lossless position meter without homodyne detection}

The best performance is achieved if intrinsic losses are negligible, $\delta_1
\ll \delta_e$, which we shall idealize as $\delta_1 = 0$.  Then,
if no homodyne detection is used (i.e., if $\Phi = \pi/2$, corresponding to
measuring the signal force as a phase modulation on the output voltage's
carrier), Eq.\ (\ref{xi_WB_lossy_pos}) predicts that $\xi_{\rm WB} \ge 1$, with
the minimum value $\xi_{\rm min}=1$ obtained for the optimal power
\begin{equation}
W_{\rm SQL} = {md^2 (\omega^2 + \delta_e^2)\omega^2 \over 2\omega_e}\;.
\end{equation}
Thus, as is well known, this conventional parametric transducer can reach but
not beat the wide-band SQL.

\subsection{Lossless position meter with ordinary homodyne detection}

By performing homodyne detection ($\Phi \ne \pi/2$), we introduce a correlation
between the position noise $x_{\rm m}(t)$ and back-action noise $F_{\rm BA}(t)$.
This correlation can be used to make the position meter perform a narrow-band
measurement of the signal force at, but {\it not} below, the narrow-band SQL,
in precisely the manner described by Eqs.\ (\ref{Snet})--(\ref{NB_SQL}) with
$\omega_0=0$.  Contrast this with the speedmeter (which, like this position
meter, uses standard homodyne detection with constant homodyne phase).
When monitoring a classical force $F_s(t)$ in a narrow-band mode, the
speed meter has net noise $S(\omega) = A'(\omega^2-\omega_{\rm meter}^2)^4 + B'$
[Eqs.\ (\ref{SAprimeBprime}) and (\ref{AprimeBprime})] and beats the
narrow-band SQL.  The
position meter has $S(\omega) = A(\omega^2-\omega_{\rm meter}^2)^2 + B$, with $AB =
\hbar^2 m^2/4$, and reaches but does not beat the narrow-band SQL.

It will be useful to reexpress this position-meter performance with constant
homodyne angle $\Phi$ in the language of $\xi_{\rm WB}(\omega)$ [Eq.\
(\ref{xi_WB_lossy_pos})].  We adjust $\Phi$ so as to minimize
$\xi_{\rm WB}(\omega)$ at some desired optimal operating frequency $\omega_{\rm
opt}$,
\begin{equation}
\cot\Phi = - {\Lambda^4\over
\omega_{\rm opt}^2(\omega_{\rm opt}^2 + \delta_e^2)}\;,
\label{omega_opt}
\end{equation}
and thereby obtain for $\xi_{\rm min} \equiv \xi_{\rm WB}(\omega_{\rm opt})$
\begin{equation}
\xi_{\rm min}^2 = - {1\over2\cot\Phi} = {W_{\rm SQL}\over W} \ll1\;.
\label{xi_min_pos}
\end{equation}
Here
$W=md^2 \Lambda^4/2\omega_e$ is the pump power [Eq.\ (\ref{Lambda})], and
Eqs.\ (\ref{omega_opt}) and (\ref{xi_min_pos}) imply that the power required
to beat the broad-band SQL is
\begin{equation}
W_{\rm SQL} = {md^2 \omega_{\rm opt}^2 (\omega_{\rm opt}^2 + \delta_e^2)\over
2\omega_e}\;.
\label{WSQL_pos}
\end{equation}
The band $\omega_1 < \omega < \omega_2$
over which $\xi_{\rm WB}^2 \le 2 \xi_{\rm min}^2$, as computed from
Eqs.\ (\ref{xi_WB_lossy_pos}), (\ref{omega_opt}) and (\ref{xi_min_pos}),
is given by
\begin{equation}
\omega_{1,2}^2 = \omega_{\rm opt}^2\left[ 1
\mp {2(\omega_{\rm opt}^2 + \delta_e^2) \over
2\omega_{\rm opt}^2+\delta_e^2} \xi_{\rm min}^2 \right ] \;.
\end{equation}

Let us compare this lossless position-meter performance with
the lossless speed meter.  Both can beat the wide-band SQL near their optimal
frequencies and they do so with approximately the same pump power [Eqs.\
(\ref{xi_min}) and (\ref{WSQL}) for speed meter, with $\delta_e^2 \sim
\omega_0^2$; Eqs.\ (\ref{xi_min_pos}) and
(\ref{WSQL_pos}) for position meter].  However, the speed meter can do so over
a wide frequency band $\omega_2-\omega_1 \agt \omega_0$
[Eqs.\ (\ref{omega12}), (\ref{xi_omega})
and associated discussion], while the position meter can only do so over a
band $\omega_2 - \omega_1 \sim \omega_{\rm opt} \xi_{\rm min}^2$ that becomes more
and more narrow as $\xi_{\rm min}$ is made smaller and smaller.  This difference is
illustrated in Fig.\ \ref{fig:xiWB},
which shows $\xi_{\rm WB}^2(\omega)$ for the two meters
with the same choice of parameters: $\xi_{\rm min}^2 = 0.1$,
optimal frequencies $\omega_0 = \omega_{\rm opt} = 1000 {\rm s}^{-1}$, and
$\delta_e = 2 \omega_0 = 2 \omega_{\rm opt} = 2000 {\rm s}^{-1}$.  (For these
parameters, the pump power $W = W_{\rm SQL}/\xi_{\rm min}^2$ is $5/4 = 1.25$ times
larger for the position meter than for the speed meter.)  The figure shows
explicitly the excellent wide-band performance of the speed meter, and
the inability of the position meter to achieve wide-band performance for this
moderately small $\xi_{\rm min} = 1/\sqrt{10} \sim 1/3$.

\begin{figure}
\epsfxsize=3.2in\epsfbox{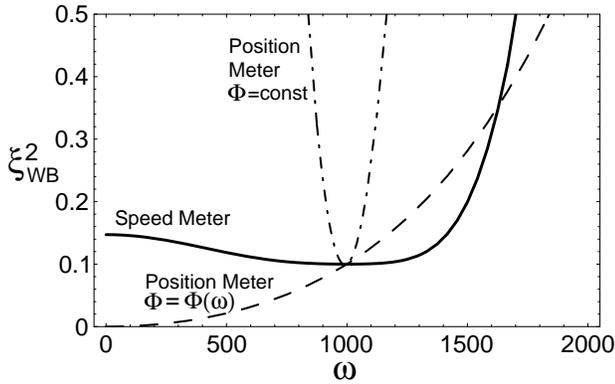}
\caption{$\xi_{\rm WB}^2(\omega)$, the squared amount by which a meter
beats the wide-band SQL when monitoring a signal force $F_s(t)$, as a
function of angular frequency $\omega$, for three meters with negligible
intrinsic losses: The speed meter
[Eqs.\ (\protect\ref{xi_omega})--(\protect\ref{WSQL})], the
position meter with homodyne detection at constant (frequency-independent)
homodyne phase $\Phi = {\rm const}$
[Eqs.\ (\protect\ref{xi_WB_lossy_pos}), (\protect\ref{omega_opt}),
(\protect\ref{xi_min_pos}) with $\delta^*=\delta_e$], and the position
meter with
optimized frequency-dependent homodyne phase $\Phi(\omega)$
(``quantum variational technique'') [Eqs.\ (\protect\ref{PhiOpt}),
(\protect\ref{xiWB_pos_var}) with $\delta_1 = 0$, $\delta^*=\delta_e$].
The parameters
of the three meters are adjusted to be the same: the same $\xi^2_{\rm min} = 0.1$
at the optimal frequency of operation $\omega_0 = \omega_{\rm opt} = 1000 {\rm
s}^{-1}$, and the same rate of extraction of signal from the resonator,
$\delta_e = 2 \omega_0 = 2 \omega_{\rm opt} = 2000 {\rm s}^{-1}$.
\label{fig:xiWB}
}
\end{figure}

\subsection{Position meter with optimized frequency-dependent
homodyne detection (``Quantum Variational technique'')}

Vyatchanin and colleagues \cite{vyat} have shown that a position meter
can be made to beat the wide-band SQL over a wide range of frequencies
by performing an (unconventional) homodyne detection with an optimized,
frequency-dependent homodyne phase $\Phi(\omega)$; they have called
this the ``quantum variational technique''.  Recently, Kimble and
colleagues \cite{kimble} have proposed a possibly practical method to
achieve such
a $\Phi(\omega)$: pass the meter's output field through a sufficiently lossless
filter that has an appropriate frequency dependence, and then perform
conventional homodyne detection.

For the above position meter, the desired, optimal frequency dependence
of the homodyne phase is the $\Phi(\omega)$ that minimizes $\xi_{\rm WB}^2(\omega)$
[Eq.\ (\ref{xi_WB_lossy_pos})]:
\begin{equation}
\cot\Phi(\omega) = - {\Lambda^4\over\omega^2(\omega^2+{\delta^*}^2)}\;,
\label{PhiOpt}
\end{equation}
where we now allow the meter to have intrinsic losses, so $\delta^* = \delta_e
+ \delta_1$.
In the idealized case that this $\Phi(\omega)$ is achieved perfectly,
the resulting performance [Eq.\ (\ref{xi_WB_lossy_pos})] is
\begin{equation}
\xi_{\rm WB}^2(\omega) = {\omega^2(\omega^2+{\delta^*}^2)\over 2\Lambda^4} +
{\Lambda^4\over 2\omega^2(\omega^2 + {\delta^*}^2)}{\delta_1\over\delta_e}\;.
\label{xiWB_pos_var}
\end{equation}

If the meter is lossless ($\delta_1 = 0$) and is adjusted to have $\xi_{\rm WB}^2 =
0.1$ at frequency $\omega = 1000 {\rm s}^{-1}$, then $\xi_{\rm WB}^2(\omega)$ has
the form shown as the dashed line in Fig.\ \ref{fig:xiWB}.  Note that
switching from constant $\Phi$ to optimized $\Phi(\omega)$ has made the
position meter broad band, though its performance above $1000 {\rm s}^{-1}$ is
not quite as good as that of the (constant-$\Phi$) speed meter.  The
pump power needed
to achieve this performance is the same (\ref{WSQL_pos}) as for the
constant-$\Phi$ position meter and nearly the same as for the speed meter.

Intrinsic losses ($\delta_1 \simeq \delta_2 \ll \delta_e$)
in the meters' resonators debilitate their low-frequency
performances
[Eq.\ (\ref{xiWB_pos_var}) for position meter; Eq.\
(\ref{xiWB_lossy}) for speed meter].
For the position meter with such losses, the minimum achievable
$\xi_{\rm WB}$ is
\begin{equation}
\xi_{\rm min} = \left({\delta_1/\delta_e}\right)^{1/4}\;.
\end{equation}
This is $\sqrt 2$ lower than for the speed meter [Eq.\ (\ref{xires3})]
at fixed $\delta_1/\delta_e$ --- though this factor $\sqrt2$ is not signficant
compared to ill-understood differences in the difficulty of realizing the two
meters. In both cases the $1/4$ power dependence on dissipation presents
serious problems for a practical device; see Sec.\ \ref{sec:realization}.

We note in passing that one can enlarge the bandwidth of the speed meter by
changing its homodyne phase from an optimized constant $\Phi$ to an optimized
frequency-dependent $\Phi(\omega)$ (analog of the above position meter).
However, the speed meter already does so well with constant $\Phi$, that the
improvement is modest. For $\xi_{\rm min}^2 = 0.1$, switching to $\Phi(\omega)$
increases the bandwidth by
about 50 per cent.  More generally, the bandwidth is widened, by switching from
constant $\Phi$ to optimized $\Phi(\omega)$, by about the same amount as it is
widened by increasing $\delta_e$ (at constant $\Phi$) by a
factor $1/\sqrt{\xi_{\rm min}}$.

\section{Speed-Meter-Based Gravitational-Wave Antenna}
\label{app:GWAntenna}

In the Laser Interferometer Gravitational-wave Observatory (LIGO), the
second generation antennas (``LIGO-II''; 2004--2007) are expected to
have sensitivites near their wide-band
SQL at $\omega \sim 2\pi \times 100$Hz \cite{lsc}.  Our speed meter research is
motivated, in part, by the goal of conceiving practical designs for a third
generation of LIGO antennas (LIGO-III) that will beat the wide-band SQL and
go into operation in
ca.\ 2008.  One possibility is the use of a microwave-based speed meter
as an internal readout device in a radically redesigned antenna (one based on
the concept of an ``optical bar'' \cite{optbar} or ``symphotonic states''
\cite{symph} or something similar).
Another possibility is an adaptation of the speed meter
into the optical band, as sketched in Fig.\  \ref{fig:4CavityIFO}.
Further possibilities will be discussed in Ref.\ \cite{kimble}.

\begin{figure}
\epsfxsize=3.2in\epsfbox{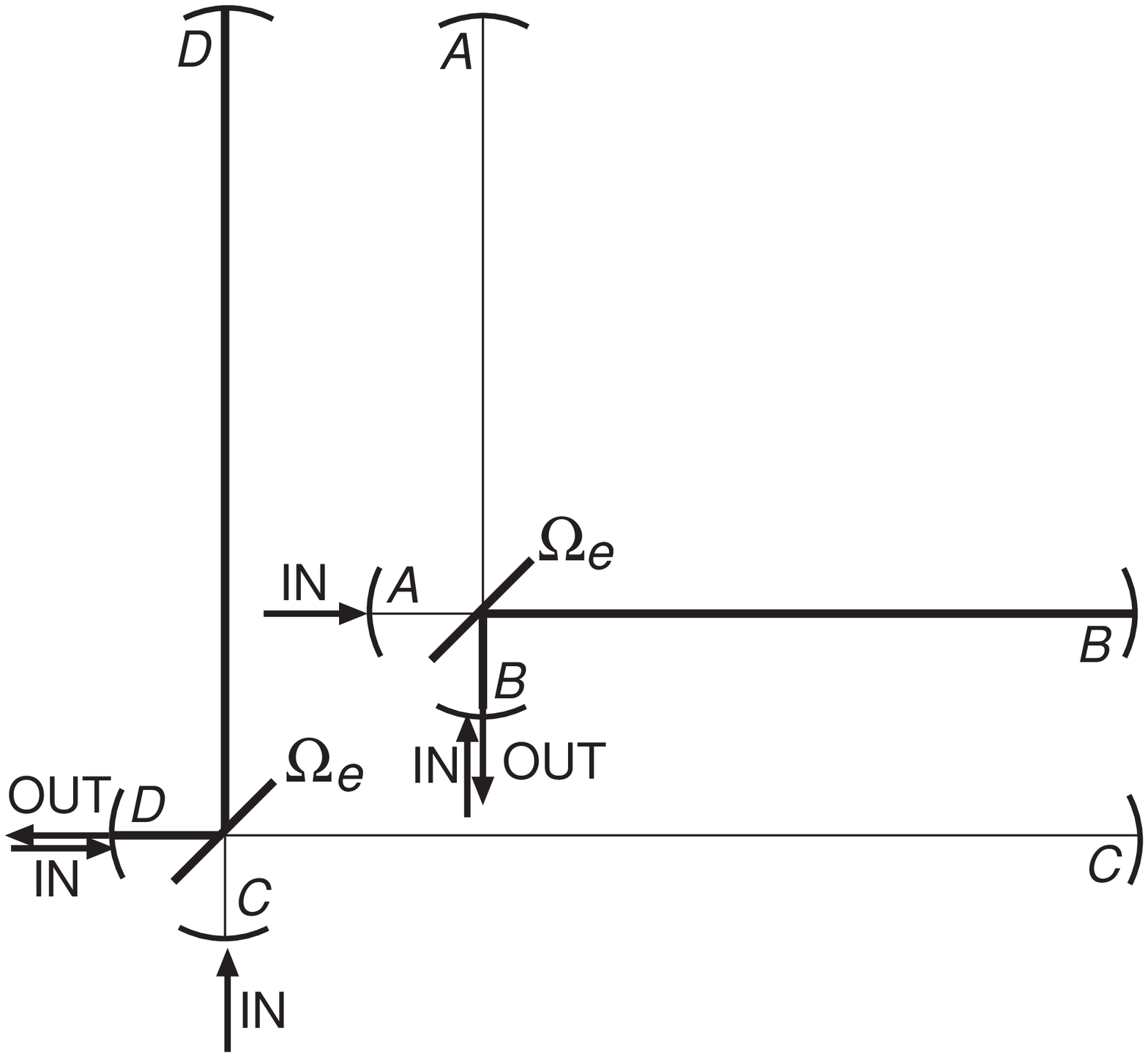}
\caption{Rough sketch of a possible LIGO antenna based on an optical-frequency
adaptation of the speed meter.
\label{fig:4CavityIFO}
}
\end{figure}

Figure \ref{fig:4CavityIFO} shows two nearly identical devices, one labeled
$11'22'$, the other labeled $33'44'$.  For the moment ignore $33'44'$.

Device $11'22'$ consists of two optical cavities (resonators)
$11'$ and $22'$ that operate
at identical resonant frequencies $\omega_e$ and are weakly coupled by
a mirror with low transmissivity.  The mirror causes light to slosh between
the two cavities with a sloshing frequency
$\Omega = c\sqrt{T}/2d$ where $T$ is the coupling
mirror's very small power
transmission coefficient and $d$ is the length of each
cavity's arm.  These cavities are the
resonators of a speed meter and $d = 4$km is the speed meter's tuning
length.  By contrast with the microwave speed meter of
Fig.\
\ref{fig:speedmeter}, which has only one test mass (that coupled to
resonator 1),
this optical speed meter has (in effect) two test masses, one coupled to
each resonator.  The reason
is that, in order to keep both resonators highly stable, all four mirrors must
be suspended as pendula, and the relative displacement $x_1$ of mirrors
$1'$ and $1$ then behaves as a test mass coupled to resonator 1, while the
relative displacement $x_2$ of $2'$ and $2$ behaves as a test mass
coupled to resonator 2.

As for the microwave speed meter, we shall read out the classical force
(gravity-wave signal) from resonator 1.  To guarantee that resonator 1
contains only a velocity signal $dx_1/dt$
[or, more precisely, a signal
that involves only $dx_1/dt$ and its time derivatives]
and not any position signal
$x_2(t)$], it is essential that resonator 2 be unexcited.
To achieve this requires, in contrast with the microwave speed meter,
that both cavities be driven by input light beams and that the relative
amplitudes
and phases of those beams be chosen appropriately.
Because resonator 2 is unexcited, its mirror motions produce
no gravity wave signal, so it does not matter whether it
is placed in the same arm as resonator 1, or in the other arm
(cf.\ Fig.\ \ref{fig:4CavityIFO}).

For the configuration in Fig.\ \ref{fig:4CavityIFO}, the two cavities
are driven by beams entering their corner mirrors.  The end
mirrors $1'$ and $2'$ have the highest possible power
reflectivities and the corner
mirrors $1$ and $2$ have more modest power reflectivities $R$ designed to
produce identical amplitude decay rates $\delta_e = c(1-R)/4d$.

As for a conventional LIGO interferometer, so for this speed meter, there
is a serious issue of frequency instability for the input light beams.
To protect against frequency fluctuations, one could proceed as in a
conventional interferometer:  Construct two identical speed meters, $11'22'$
and $33'44'$ as shown in Fig.\ \ref{fig:4CavityIFO}, with the strongly
excited resonators 1 and 3 in the two orthogonal arms of the LIGO vacuum
system.  Drive the four cavities with phase coherent light beams that are
all phase
locked to the same master oscillator. Construct the difference of
the outputs from 1 and 3 by mixing at a beam splitter, and perform
the homodyne detection on that difference.  As for a conventional
interferometer, such a scheme should provide significant protection against
frequency fluctuations.

Although we have not yet carried out a full and detailed analysis of this
optical speedmeter, our approximate analyses show that, up to factors of
order unity, its performance is described by the same equations as
for the microwave speedmeter.
It can beat the wide-band SQL by the factors $\xi_{\rm WB}(\omega)$
derived and discussed in Secs.\ \ref{sec:WB_lossless_sensitivity},
\ref{sec:NB_lossless_sensitivity} and \ref{sec:losses}.

More specifically,
if such an optical speed meter is optimized as in Sec.\ \ref{sec:losses}
$(\delta_e \simeq 2\omega_0'$, $\Omega \simeq \sqrt{3} \omega_0'$ where
$\omega_0'$ is the optimal frequency of operation),
then to reach the wide-band SQL at $\omega =
\omega_0'$ requires a pump power
\begin{equation}
W = W_{\rm SQL} \simeq
4md^2{\omega_0'}^4/\omega_e
\label{power'}
\end{equation}
[Eq.\ (\ref{power})], and by using a pump power $W$ that exceeds this $W_{\rm
SQL}$ and achieving sufficiently low optical losses $\delta_1 \ll \delta_e$,
the wide-band SQL can be beat in the vicinity of the optimal frequency
$\omega_0'$ by a factor
\begin{equation}
\xi_{\rm min} = \sqrt{W_{\rm SQL}\over W}\;; \quad
\xi_{\rm min} \agt \left({4\delta_1\over\delta_e}\right)^{1/4}
\label{ximin'}
\end{equation}
[Eqs.\ (\ref{power}) and (\ref{xires3})].

Note that the SQL power $W_{\rm SQL}$ corresponds to a
stored energy in each resonator $11'$ and $33'$ given by
\begin{equation}
{\cal E}_{\rm SQL} = {W_{\rm SQL}\over 2\delta_e} \simeq {md^2{\omega_0'}^3
\over\omega_e}\;.
\end{equation}
This is the same stored energy (to within a factor of order unity)
as is required to reach the SQL
in a conventional LIGO gravitational-wave detector \cite{symph}.
This stored energy and the corresponding circulating light power $W^{\rm
circ}_{\rm SQL}$ in the resonators are uncomfortably large:
\begin{equation}
W^{\rm circ}_{\rm SQL} = {{\cal E}_{\rm SQL}\over 2d/c} \sim 900{\rm kW}\;,
\label{Wcirc}
\end{equation}
where we have used $m=11$kg, $d=4$km, $\omega_0' = 2\pi\times 100$Hz,
and $\omega_e = 1.8 \times 10^{15} {\rm s}^{-1}$ (wavelength
$1.06\mu$m), as planned for LIGO \cite{lsc}.
There is hope, in LIGO, of operating at circulating powers of this order
\cite{lsc}, but to do so will be extremely challenging.  And to beat the
SQL by a factor $\xi_{\rm min}$ at the optimal frequency $\omega_o'$ using
the optical speed meter of Fig.\ \ref{fig:4CavityIFO} would require
an even larger circulating power
\begin{equation}
 W^{\rm circ} = W_{\rm SQL}^{\rm circ} /\xi_{\rm min}^2\;
\end{equation}
[Eqs.\ (\ref{ximin'})--(\ref{Wcirc})].
Moreover, even if such extreme power could be handled in LIGO-III, the
resonators' optical losses might not be much smaller than
$\delta_1/\delta_e \sim 0.01$, which corresponds to a limit on the
achievable sensitivity $\xi_{\rm min} \agt
(4\delta_1/\delta_e)^{1/4} \simeq 0.4$
(and an increase in event rate for gravitational-wave bursts of
$\alt 1/0.4^3 \simeq
15$ over an SQL-limited interferometer).

Although this scheme is rather complex and places extreme demands on the
circulating light power and on optical losses, it nevertheless might turn
out to be practical.  Moreover, it is not significantly more
complex or demanding than schemes that have been devised
for beating the SQL in LIGO-III
by modifying a conventional interferometer's input
and/or output optics \cite{unruh,vyat,kimble}.

The high power demands of all these schemes leave our
research groups dissatisfied and motivate our continuing efforts to find
more promising designs that entail much less optical power---schemes
that might resemble those described in Refs.\
\cite{optbar,symph}.


\begin{references}

  \bibitem{speedmeter} V.\ B.\ Braginsky and F.\ Ya.\ Khalili, Phys.\  Lett.\
  A {\bf 147}, 251 (1990).

  \bibitem{QuantumMeasurement} V.\ B.\ Braginsky and F.\ Ya.\ Khalili, {\it
  Quantum Measurement} (Cambridge University Press, Cambridge, 1992).

  \bibitem{NL96} V.\ B.\ Braginsky and F.\ Ya.\ Khalili, Phys.\  Lett.\
  A {\bf 218}, 167 (1996).

  \bibitem{optbar} V.\ B.\ Braginsky, M.\ L.\ Gorodetsky and F.\ Ya.\ Khalili,
  Phys.\  Lett.\ A {\bf 232}, 340 (1997).

  \bibitem{symph} V.\ B.\ Braginsky, M.\ L.\ Gorodetsky and F.\ Ya.\
  Khalili, Phys.\  Lett.\ A {\bf 246}, 485 (1998).

  \bibitem{yurke} These equations should be fairly obvious from the classical
    theory of microwave networks.
    For foundations that justify them as quantum mechanical Heisenberg-Picture
    equations see, e.\ g., B.\ Yurke and J.\ S.\ Denker, Phys.\ Rev.\ A
    {\bf 29}, 1419 (1984).

  \bibitem{syrtsev_khalili} A.\ V.\ Syrtsev and F.\ Ya.\ Khalili, JETP, {\bf
    79}, 409 (1994).

  \bibitem{Mitrofanov} V.\ B.\ Braginsky, V.\ P.\ Mitrofanov and K.\ V.\
  Tokmakov, Phys.\ Lett.\  A {\bf 218}, 164 (1996) and unpublished
  subsequent research.

  \bibitem{Whisp} V.\ B.\ Braginsky, V.\ S.\ Ilchenko and K.\ S.\ Bagdasarov,
  Phys.\ Lett.\ A, {\bf 120}, 300 (1987).

  \bibitem{SSD} V.\ B.\ Braginsky and V.\ P.\ Mitrofanov, {\it Systems with
  Small Dissipation} (University of Chicago Press, Chicago, 1985).

  \bibitem{MW_resonator}
  I.\ A.\ Bilenko, E.\ N.\ Ivanov, M.\ E.\ Tobar, D.\ G.\ Blair, Phys.\
  Lett.\ A {\bf 211}, 136 (1996).

  \bibitem{cuthbertson}
  B.\ D.\ Cuthbertson, M.\ E.\ Tobar, E.\ I.\ Evanov and D.\ G.\ Blair,
  IEEE Trans.\ Ultrason.\ Ferroelect.\ Freq.\ Control, {\bf 45}, 1303 (1998).

  \bibitem{Blair} T.\ Suzuki, P.\ Turner, J.\ Ferreirinho, D.\ G.\ Blair and
  R.\ S.\ Crisp, J.\ Low Temp.\ Phys., {\bf 58}, 37 (1985).

  \bibitem{vyat}
  S.\ P.\ Vyatchanin and A.\ B.\ Matsko, JETP {\bf 77}, 218 (1993);
  S.\ P.\ Vyatchanin and E.\ A.\ Zubova, Phys.\ Lett.\ A {\bf 201}, 269
  (1995);
  S.\ P.\ Vyatchanin and A.\ B.\ Matsko, JETP {\bf 82}, 1007 (1996).
  S.\ P.\ Vyatchanin, Phys.\ Lett.\ A, {\bf 239}, 201 (1998).

  \bibitem{kimble} H.\ J.\ Kimble, Yu.\ Levin, A.\ B.\ Matsko, K.\ S.\ Thorne
  and S.\ P.\ Vyatchanin, paper in preparation.

  \bibitem{lsc} R.\ Weiss et.\ al., {\it LSC White Paper on Detector Research
  and Development} (LIGO Project Document XXXXX, November 1998).

  \bibitem{unruh} W.\ G.\ Unruh, in {\it Quantum Optics, Experimental
  Gravitation, and Measurement Theory}, eds. P.\ Meystre and M.\  O.\
  Scully, (Plenum, 1982), p.\  647.

\end{references}
\end{document}